\documentclass[a4paper,11pt]{article}

\pdfoutput=1
\usepackage{jcappub}

\usepackage{graphicx}
\usepackage{longtable}
\usepackage{float}
\usepackage{dcolumn}
\usepackage{bm}
\usepackage{appendix}
\usepackage{multirow}
\usepackage{color}
\usepackage[utf8]{inputenc}
\usepackage{footmisc}
\usepackage{hyperref}

\usepackage{xcolor}
\usepackage{graphicx}
\usepackage{bm}
\usepackage{ulem}
\usepackage{multirow, array, float, tabularx}
\usepackage{amsmath}
\hypersetup{
	citecolor = red,
	linkcolor = blue,
	urlcolor = magenta
}

\newcommand{\be}{\begin{equation}}
	\newcommand{\ee}{\end{equation}}
\newcommand{\bea}{\begin{eqnarray}}
	\newcommand{\eea}{\end{eqnarray}}
\newcommand{\bes}{\begin{subequations}}
	\newcommand{\ees}{\end{subequations}}

\newcommand{\bc}{\begin{center}}
	\newcommand{\ec}{\end{center}}

\usepackage{wasysym}

\begin{document}

    \title{A comparative analysis of dissipation coefficients in warm inflation}

    \author[a]{F. B. M. dos Santos}\emailAdd{fbmsantos@on.br}
    \author[a]{R. de Souza} \emailAdd{rayffsouza@on.br}
    \author[a]{J. S. Alcaniz}\emailAdd{alcaniz@on.br}

\affiliation[a]{Departamento de Astronomia, Observatório Nacional, 20921-400, Rio de Janeiro - RJ, Brazil}

\abstract{In the warm inflation scenario, the early cosmic acceleration is driven by the inflaton coupled to thermal fields,  decaying into radiation and leaving a hot universe populated by relativistic particles after the end of inflation. The interaction
is usually modeled by a dissipation coefficient $\Upsilon$ that contains the microphysics of the model. In this work, we adopt a well-motivated potential $V(\phi)=\frac{\lambda}{4}\phi^4$ and constrain a variety of $\Upsilon$ parameterizations by using updated  Cosmic Microwave Background data from the  \textit{Planck} and \textit{BICEP/Keck Array} collaborations. We also use a Bayesian statistical criterion to compare the observational viability of these models. Our results show a significant improvement in the constraints over past results reported in the literature and also that some of these warm inflation models can be competitive compared to Starobinsky inflation.}

	\maketitle

\section{Introduction}\label{sec1}

The cosmological data obtained by recent Cosmic Microwave Background (CMB) experiments  \cite{WMAP:2012nax,Planck:2018nkj,Planck:2018vyg,BICEP:2021xfz} can give us a comprehensive understanding of the cosmic history from very early times. They can place tight constraints on the amplitude of the scalar primordial spectrum, being almost scale invariant at more than $2\sigma$. Such constraints provide a good indication that an inflationary period took place in the early universe, and that the most compelling theoretical explanation is that it is driven by the presence of a scalar field $\phi$ dubbed inflaton \cite{Guth:1980zm,Starobinsky:1980te,Linde:1981mu}, which follows the dynamics dictated by a potential $V(\phi)$. It is then possible to place significant limits on and test the physical viability of the many inflationary models proposed so far; simple models, such as the one given by the monomial potential, are very disfavored by data, as the parameter space of the inflationary parameters excludes some of these models if the simplest canonical picture is considered.

Although other additional ingredients can be considered in order to restore the concordance of these models with data (see, e.g., \cite{Campista:2017ovq} and references therein), one attractive idea involves the realization of a reheating process during inflation. Due to interactions of the inflaton with thermal fields, dissipation of the field energy into radiation can happen concomitantly with the slow-roll regime that realizes the accelerated expansion. In this scenario, called \textit{warm inflation} (WI) \cite{Berera:1995ie,Maia:1999yt,Hall:2003zp,Berera:2023liv}, the universe would smoothly go into the radiation era at the end of inflation, with little to no need for an additional preheating/reheating period, whose dominant mechanism is still unknown  and challenging to be probed by current observational data. Also, the warm inflation picture provides a rich phenomenology in that the cosmological observables can radically differ from what their cold inflation counterparts predict. In particular, the tensor-to-scalar ratio $r$ can be very low, of order $10^{-4}$ when WI happens in the weak dissipative regime \cite{Benetti:2016jhf}, while a very strong regime gives a striking result of $r\sim 10^{-29}$ \cite{Das:2020xmh,Kamali:2019xnt}, potentially excluding the presence of a stochastic gravitational wave background produced during inflation within these models. Important physical aspects in the WI framework, including the $\eta$-problem and the swampland conjectures, have been discussed in the recent literature, see e.g. \cite{Baumann:2014nda,Berera:2008ar,Bastero-Gil:2019gao,Das:2018hqy,Das:2018rpg,Das:2019acf,Kamali:2019xnt,Berera:2019zdd,Das:2020xmh,Santos:2022exm}.

While the phenomenology of warm inflation is well established in the literature, the observational viability of this class of models, in light of the available CMB data, still needs to be investigated more. Even if a particular WI model provides a good description of  the current observations, it is still being determined whether it is favored or disfavored over cold inflation scenarios from the statistical point of view. Recent works have started to investigate this issue \cite{Benetti:2016jhf,Bastero-Gil:2017wwl,Arya:2017zlb,Benetti:2019kgw}; for instance, in \cite{Arya:2018sgw}, a  full numerical analysis was carried out to study models characterized by quartic and sextic power-law potentials. The results show consistency with the allowed region in $n_s - r$ plane, where $n_s$ is the spectral index. 

This work investigates which behavior of the dissipation coefficient $\Upsilon$ that realizes the energy conversion process is preferred when the full \textit{Planck} CMB data is considered. Through a Bayesian selection analysis, we also compare the theoretical predictions of the corresponding WI model with the well-studied Starobinsky inflation. In our analysis, we assume the quartic potential $V(\phi)=\lambda\phi^4/4$ and consider four forms for $\Upsilon$, characterizing the most typical models studied in the literature, with motivations that range from concrete particle physics backgrounds to more phenomenological ones \cite{Motaharfar:2018zyb}. Some $\Upsilon$ forms are well-motivated within the standard particle physics formalism, such as the one proposed in \cite{Bastero-Gil:2016qru}, in which the dissipation coefficient is given by $\Upsilon\propto T$, with $T$ being the temperature of the thermal bath. 

Our objective here is twofold: first, we shall obtain updated constraints on the models investigated in \cite{Bastero-Gil:2017wwl,Arya:2017zlb,Arya:2018sgw} by considering the most recent \textit{Planck} 2018+lensing data, as well as the impact of CMB B-mode polarizations, as estimated by the latest release of the \textit{BICEP/Keck Array} observatories \cite{BICEP:2021xfz}. This allows us to obtain additional constraints on the inflation scenarios to determine each model's dissipation range. Second, we employ a Bayesian statistical criterion to determine the preferred dissipation form by data, guiding our model building within warm inflation and our approach to future data. The results of the model selection also allow us to compare the observational viability of the WI scenarios with the well-established Starobinsky model, providing a clear benchmark for our findings.

The work is organized as follows. In sec. \ref{sec2} we summarize the idea of warm inflation, while in Sec. \ref{sec3} we briefly discuss the models investigated. Sec. \ref{sec4} is devoted to the methods used to solve each model and to implement them into the numerical code. In Sec. \ref{sec5} we describe the data sets used in the statistical analysis, while Sec. \ref{sec6} presents our results. Finally, in Sec. \ref{sec7}, we present our considerations and perspectives for future works.

\section{Warm inflation}\label{sec2}

During the slow-roll regime, the inflaton field is coupled to degrees of freedom that drive the production of radiation through dissipation. This can be represented by a dissipation coefficient $\Upsilon$ present in the equations of motion, characterizing an interaction between fields as
\begin{gather}
\ddot\phi + 3H\dot\phi + V_{,\phi} = -\Upsilon\dot\phi\\
\dot\rho_r + 4H\rho_r = \Upsilon\dot\phi^2\\
3H^2M_p^2 = \frac{\dot\phi^2}{2} + V(\phi) + \rho_r
\end{gather}
where $M_p\equiv\frac{1}{\sqrt{8\pi G}}$ is the reduced Planck mass, while $H=\frac{\dot a}{a}$ is the Hubble parameter, with a dot denoting a derivative with respect to cosmic time. The slow-roll regime is determined in the same manner as in the usual cold inflation, with the difference that, due to the presence of the dissipative term, the slow-roll parameters are redefined as
\begin{equation}
\epsilon_w = \frac{\epsilon_V}{1+Q}, \quad \eta_w = \frac{\eta_V}{1+Q},
\end{equation}
with $\epsilon_V$ and $\eta_V$ being the usual cold inflation slow-roll parameters, $\epsilon_V\equiv M_p^2(V_{,\phi}/V)^2/2$, and $\eta_V\equiv M_p^2 V_{,\phi\phi}/V$ respectively, while $Q$ represents the ratio between the rates of dissipation and the expansion of the universe
\begin{equation}
Q\equiv\frac{\Upsilon}{3H}.
\end{equation}
This is an important quantity, since it characterizes the regime in which warm inflation will happen. For $Q\ll 1$, we have the \textit{weak} dissipative regime, while the strong regime is realized by $Q\gg 1$. Ultimately, the regime of dissipation that is required to fit CMB observations is dependent on both the dissipation coefficient and the choice of the inflationary potential \cite{Bastero-Gil:2017wwl,Arya:2017zlb,Montefalcone:2022jfw}.

Furthermore, the slow-roll approximations in the warm inflation picture generally assume that thermalization of the radiation bath is achieved rapidly, such that its energy density is almost constant, and that the potential energy still dominates the energy content of the universe. As a result, the ratio of the temperature $T$ and $H$ must obey $T/H>1$ in order for warm inflation to be characterized. Also, we have defined $\tilde g = \frac{\pi^2 g_\star}{30}$, with $g_\star = 106.75$ being the number of relativistic degrees of freedom in the early universe\footnote{We expect some constructions of WI to require distinct values for the relativistic degrees of freedom. Nevertheless, in our approach to the models, the results have a negligible dependence on this parameter.} \cite{Baumann:2022mni}. 
The equations of motion in the slow-roll regime will then become
\begin{gather}
3H(1+Q)\dot\phi + V_{,\phi} \simeq 0, \\
\rho_r \simeq \tilde g T^4, \\
H^2\simeq\frac{V}{3M_p^2}.
\end{gather}
The inflaton energy density is still dominating during warm inflation, but as the process goes on, the dissipation of the field that leads to the increase in temperature will also increase the amount of radiation energy density, so that when inflation ends, depending on the dissipation regime, the universe might be in a state of radiation domination. As a result of dissipative effects, the observational quantities are expected to change as well. The primordial power spectrum is modified with the addition of extra terms as \cite{Graham:2009bf,Bastero-Gil:2009sdq,Bastero-Gil:2011rva}

\begin{equation} \label{9}
\Delta^2_{\mathcal{R}}(k_\star)  = \mathcal{P}_{\mathcal{R},c}\left(1 + 2n_{BE,\star} + \frac{2\sqrt{3}\pi Q_\star}{\sqrt{3+4\pi Q_\star}}\frac{T_\star}{H_\star}\right)G(Q_\star)
\end{equation}
where $\mathcal{P}_{\mathcal{R},c} \equiv \left(\frac{H_\star^2}{2\pi\dot\phi_\star}\right)^2$ and $n_{BE,\star}=\frac{1}{e^{H_\star/T_\star}+1}$, where the subscript $_\star$ corresponds to quantities measured at the CMB scale that leaves the horizon during inflation. $G(Q_\star)$ results from the coupling between inflaton and radiation perturbations, being dependent on the dissipation coefficient and the potential that is considered and determined numerically. We note that the prefactor of Eq. (\ref{9})  represents the cold inflation result, meaning that, in general, the extra terms can amplify the spectrum. Since there are no direct couplings between the tensor modes and the radiation bath, the tensor power spectrum has its usual form
\begin{equation}
\Delta^2_T(k_\star) = \frac{2H_{\star}^2}{\pi^2 M_p^2}.
\label{10}
\end{equation}
From these quantities, one can determine the well-known spectral index $n_s$ and tensor-to-scalar ratio $r$ as
\begin{equation}
n_s-1 = \frac{d\ln\Delta^2_{\mathcal{R}}}{d\ln k}, \quad r=\frac{\Delta^2_T}{\Delta^2_\mathcal{R}},
\end{equation}
constrained by the \textit{Planck} collaboration as $n_s=0.9645\pm 0.0042$, while the amplitude of the primordial spectrum is estimated as $\log(10^{10}A_s)=3.044\pm 0.014$, at the pivot scale $k_\star=0.05$ Mpc$^{-1}$, both at 68\% (C.L.). When tensor fluctuations are considered, the $\Lambda$CDM$+r$ model gives the upper limit on the tensor-to-scalar ratio as $r<0.056$ \cite{Planck:2018jri}, if \textit{Planck} data is combined with the \textit{BICEP/Keck Array} CMB B-mode polarization measurements from the 2015 observation season (BK15). On the other hand, more recent data from the 2018 observation season (BK18) imposes an even more strict upper bound on $r$, of $r<0.036$ \cite{BICEP:2021xfz}, reinforcing the smallness of primordial tensor perturbations. Consequently, these limits become crucial to restrict the parameter space of many potential models, including warm inflation ones, which generally allow for a very small tensor-to-scalar ratio.

\section{The quartic warm inflation model}\label{sec3}

It is well known that the potential given by 
\begin{equation}
V(\phi)=\frac{\lambda}{4}\phi^4,
\end{equation}
is no longer supported by current CMB data as an inflationary model, when the simplest picture of a minimally-coupled, cold inflation scenario is considered. However, when interactions of the inflaton with environmental fields are non-negligible, a significant change in the description of the model is perceived, such that the predictions at the $n_s-r$ level can be concordant with the most recent \textit{Planck} restrictions. Such interactions are often modeled by the presence of a dissipation coefficient $\Upsilon$, that describes the microphysics involved and generally depends on the couplings of the inflaton with the relevant fields in the thermal bath \cite{Bastero-Gil:2010dgy}. The temperature and field dependence of $\Upsilon$ can be expressed in a more general manner by the following expression
\begin{equation}
    \Upsilon=C_\Upsilon T^p \phi^c M^{1-p-c},
\end{equation}
for constants $C_\Upsilon$ and $M$.

Below, we briefly describe the most studied forms so far in the literature, as discussed in \cite{Kamali:2023lzq,Motaharfar:2018zyb}:
\begin{itemize}
\item $\Upsilon = C_C\frac{T^3}{\phi^2}$: This form is motivated by supersymmetric arguments. Here, the inflaton interacts with superfields that subsequently decay into radiation, in the so-called two-stage mechanism \cite{Bastero-Gil:2012akf}, resulting in the `low-temperature' regime of WI. Also, it was shown recently that WI can be realized if the inflaton has an axion-like coupling to Yang-Mills gauge fields \cite{Berghaus:2019whh}. In this case, the friction of the inflaton's motion comes from topological transitions between different vacua, which increase with temperature. This scenario renders a dissipation coefficient that scales with $T^3$ in the limit of a small inflaton mass. 
\item $\Upsilon = C_L T$: The most solid motivation for this coefficient comes from considering the inflaton as a pseudo Nambu-Goldstone boson, arising from the relative phase between two complex scalars which collectively break a $U(1)$ gauge symmetry \cite{Bastero-Gil:2016qru}. The scalars are then coupled to fermionic degrees of freedom, which all possess an interchange symmetry that protects the inflaton mass from large thermal corrections. In the high temperature limit, the dissipation coefficient takes a linear dependence on the heat bath temperature.
\item $\Upsilon = C_I\frac{\phi^2}{T}$: Similar to the mechanism of \cite{Bastero-Gil:2016qru}, the authors in \cite{Bastero-Gil:2019gao} considered a direct coupling of the symmetry breaking scalars with two other complex scalar fields, instead of fermions. As a consequence of the distinct statistical nature of the coupled fields, the dissipation coefficient acquires an inverse dependence on the temperature, when considering the leading thermal contributions to the masses of the auxiliary particles. This behavior is particularly interesting since, as showed in \cite{Bastero-Gil:2019gao}, the inflationary dynamics can occur in the strong dissipative regime ($Q \gg 1$), while producing perturbations consistent with CMB observations.
\item $\Upsilon = C_H H$: A dissipation proportional to the Hubble rate has been applied in the context of energy exchange in the universe's dark sectors (see e.g \cite{Bolotin:2013jpa,vonMarttens:2019wsc}). Also, it has recently been used in studies relating dissipation effects with effective viscosities \cite{Barbosa:2017ojt}. In the WI picture, it produces analytical expressions, since the dissipation parameter $Q$ is effectively constant, and can be regarded as a phenomenological approach \cite{Motaharfar:2018zyb}.
\end{itemize}

The general phenomenology of these models was discussed in recent literature. In \cite{Motaharfar:2018zyb}, the discussion is centered around the capability of these forms for $\Upsilon$ in light of the proposed swampland conjectures within warm inflation. In essence, the conjectures are easily satisfied if inflation takes place in the strong dissipative regime, which, in the case of the quartic potential, is not allowed for every $\Upsilon$, as in most cases the $n_s-r$ data usually indicates a preference for a weak regime.  

Conversely, a more concrete answer in terms of recent robust cosmological data is still poorly provided. Only a few works have dealt with a full CMB analysis of these models. In \cite{Arya:2017zlb,Arya:2018sgw}, the authors analyzed the quartic model for $\Upsilon=C_C\frac{T^3}{\phi^2}$ and $\Upsilon=C_L T$, with \textit{Planck} 2015 data, finding an indication for the weak dissipative regime. Similar results are found in \cite{Bastero-Gil:2017wwl}, where they  focused on the linear dissipative coefficient. However,
none of these analyses compare the observational viability of the WI scenarios with other
well-established cold inflation models.  Moreover, the recent release from CMB B-mode polarization data from the \textit{BICEP/Keck Array} telescopes has imposed severe constraints on well-known inflationary models. Therefore, we shall run an updated analysis of the models discussed previously, as well as including the inverse and $\propto H$ dissipative coefficients into this discussion, in order to estimate what order of dissipation during inflation is allowed, as well as the overall preference by the current cosmological data.

\begin{figure*}[t]
\centering
\includegraphics[width=0.6\columnwidth]{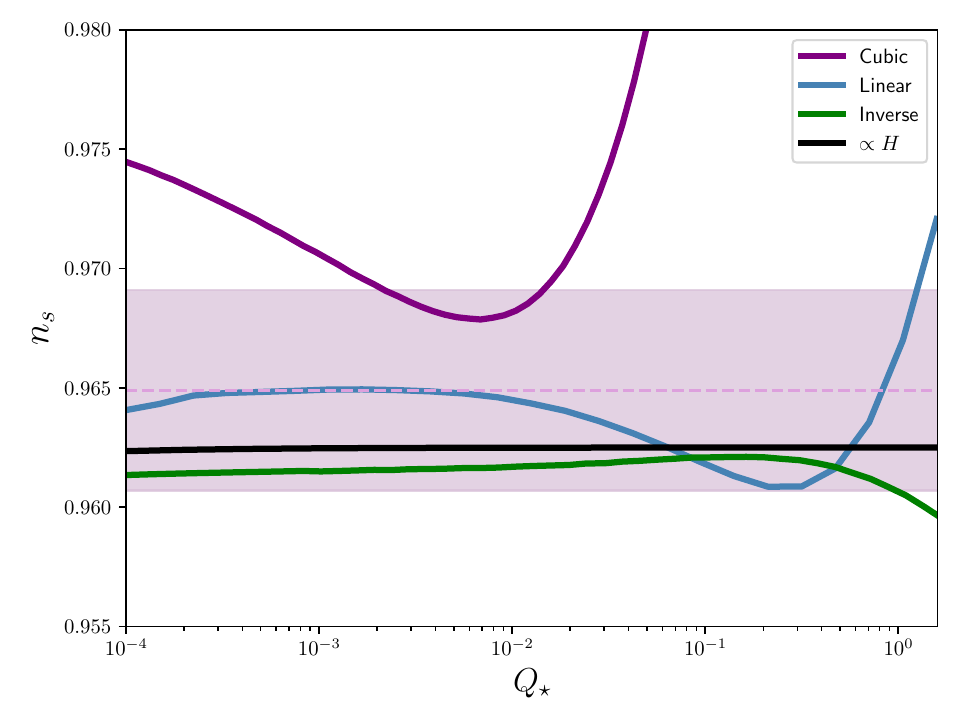}
\caption{The spectral index $n_s$ as a function of $Q_\star$ for the four models investigated in this work, namely a quartic potential with a dissipative coefficient with cubic (purple curve), linear (blue curve), inverse (green curve) dependence on the temperature, as well as the one given by $\Upsilon\propto H$ (black curve). To plot these curves, we fixed $N_\star=60$, while the horizontal band corresponds to the 68\% confidence level limits on $n_s$ imposed by \textit{Planck} data.}
\label{fig:1}
\end{figure*}

An initial estimation can be obtained by comparing the inflationary parameters with current CMB constraints. Figure \ref{fig:1} shows the spectral index $n_s$ as a function of dissipative ratio at horizon crossing $Q_\star$ for the four $\Upsilon$ considered. The curves are compared with the 68\% limit on $n_s$ obtained by \textit{Planck}. A common feature among the models is that the weak dissipative regime is in excellent agreement with the favored region, although the linear ($\Upsilon\propto T$), inverse ($\Upsilon\propto T^{-1}$) models and especially the $\Upsilon\propto H$ one allow for a initial strong regime, as $Q\sim 1$. This is a good indication that the models at hand can reproduce the data quite well, motivating a direct comparison with robust data, in order to extract more properties of each model.

\section{Methods}\label{sec4}

In order to numerically implement these models, it is necessary to know the behavior of the primordial spectra (\ref{9}) and (\ref{10}) as function of scale $k$. In the cold inflation scenario, both $\Delta^2_\mathcal{R}$ and $\Delta^2_T$ are parameterized as functions of $k$ and are given by a power-law behavior (in the absence of running) tilted by the scalar and tensor spectral indices $n_s$ and $n_t$, respectively:
\begin{equation}
\Delta^2_\mathcal{R}=A_s\left(\frac{k}{k_\star}\right)^{n_s-1}, \quad \Delta^2_T=A_t\left(\frac{k}{k_\star}\right)^{n_t}.
\end{equation} 
In the WI framework, this can be done by integrating the equation of motion for $Q$ as a function of $k$. In the slow-roll regime of the quartic potential, $\phi(Q_k)$ is easily achieved, such that $Q(k)$ becomes the only quantity we should know in order to characterize the behavior of each model. The background equation of $Q$ depends on the choice of dissipation coefficient \cite{Kamali:2023lzq} and is usually given with the number of e-folds $N = \ln a$ as the time variable. Alternatively, we can use the moment of which a given scale crossed the horizon at $k = a H$, some $N$ e-folds before the end of inflation. Therefore, using $d \ln k/ d N = 1 - \epsilon_H \approx 1 - \epsilon_w$ \cite{Das:2022ubr}, where $\epsilon_H = -\dot H/H^2$, we then solve the differential system to obtain the scale dependence of background quantities.

Thus, we parameterize $\Delta^2_\mathcal{R}(k/k_\star)$ and $\Delta^2_T(k/k_\star)$ as a function of scale - according to (\ref{9}) and (\ref{10}), respectively - through an external script. Then the Code for Anisotropies in the Microwave Background (CAMB) Boltzmann solver \cite{Lewis:1999bs,Howlett:2012mh} reads each realization of the spectrum and computes the coefficients of the CMB spectra. We choose $k=0.05$Mpc$^{-1}$ as representing the pivot scale that leaves the horizon at some e-fold number $N_\star=\ln a_\star$ before the end of inflation. To estimate $N_\star$ for each model, we follow \cite{Liddle:2003as}, with the assumption that the end of inflation is immediately followed by the radiation era, so that $\frac{a_{end}}{a_{reh}}=1$. Additionally, considering that entropy is conserved throughout cosmic history, one eventually arrives at the expression
\begin{equation}
    \frac{k_\star}{a_0H_0} = e^{-N_\star}\left(\frac{43}{11 g_\star}\right)^{1/3}\frac{T_0}{T_{end}}\frac{H_\star}{H_0}\; .
    \label{15}
\end{equation}
We include the numerical computation of $N_\star$ into our code, so the number of e-folds has now a $Q_\star$ dependence. We bring here as an example the WI scenario done in \cite{Das:2020xmh}, in which a model in the strong regime was implemented. In that setting, due to the high dissipative terms, the pivot scale leaves the horizon at around $N_\star\simeq 48$; we then expect that in the weak regime a larger $N_\star$ will be needed when we consider Eq. (\ref{15}). We also emphasize that, in our approach, $\lambda$ is also a derived parameter, given by the normalization of the scalar spectrum in (\ref{9}) at $k=k_{\star}$. Therefore it is determined by choices of $A_s$ and $Q_\star$, which are sampled parameters in our analysis.

To compare the different dissipation coefficients considered above in light of recent CMB data, we compute the Bayesian evidence for each case. Given a model $\mathcal{M}$\footnote{By model we refer to a specific form of the dissipation coefficient, while always considering the quartic potential.} with a set of parameters $\theta$, the evidence is calculated as the average of the likelihood function $\mathcal{L}(\theta)$ under the prior probability $p(\theta|\mathcal{M})$ \cite{Trotta:2017wnx,SantosdaCosta:2017ctv,Cid:2018ugy,Benetti:2017juy}:
\begin{equation}\label{evidence}
    \mathcal{Z} = \int_{\Omega_\mathcal{M}} \mathcal{L}(\theta)p(\theta|\mathcal{M})\mathrm{d}\theta\; .
\end{equation}
\indent In the above equation, the integral is performed over the parameter space $\Omega_\mathcal{M}$ for a specific model. The difference of the logarithm of the evidences is given by the Bayes' factor $\ln B_{ij} \equiv \ln \mathcal{Z}_i - \ln \mathcal{Z}_j$. Thus, a positive/negative $\ln B$ favors/disfavors the model $i$ in comparison with model $j$ \cite{doi:10.1080/01621459.1995.10476572}. A manner of comparing two different models is to use the so-called Jeffreys' scale \cite{jeffreys1998theory,Trotta:2017wnx}, in which the $\ln B_{ij}$ computed can take the values $<1$ (inconclusive), $1-2.5$ (weak), $2.5-5$ (moderate) and $>5$ (strong), as evidences for/against a proposed model, if the sign is positive/negative.

\section{Observational data}\label{sec5}

As discussed earlier, for each dissipation coefficient, the viable range of $Q_\star$ for which there is concordance with the $n_s-r$ \textit{Planck} constraints will be different. This result has a direct implication on which dissipative regime warm inflation can happen, with consequences on the particle production regime. We implement the models considered in the CAMB code and use their output to perform a statistical analysis by the Monte Carlo Markov Chain (MCMC) approach, through the \texttt{cobaya} code \cite{Torrado:2020dgo,2019ascl.soft10019T}, in which the necessary likelihoods and data sets are implemented. In particular, we take the recent TTTEEE \texttt{HiLLiPoP} high-$\ell$ likelihood \cite{Tristram:2023haj} \footnote{\url{https://github.com/planck-npipe/hillipop}}, together with the low-multipole \texttt{Commander} for TT modes \cite{Planck:2018yye,Planck:2019nip} and the \texttt{LoLLiPoP} one for EE modes \footnote{\url{https://github.com/planck-npipe/lollipop}}. Furthermore, we combine this with the PR4 likelihood for the CMB lensing potential \cite{Carron:2022eyg}. Since warm inflationary models usually predict a very low tensor-to-scalar ratio, we also consider further constraints from CMB lensed B-mode polarization data from the \textit{BICEP/Keck Array} collaboration, from the recent 2018 observation season \cite{BICEP:2021xfz}. Its combination with \textit{Planck} data has put severe limits on the $n_s-r$ confidence region (see the Fig. 5 on \cite{BICEP:2021xfz}, also \cite{Tristram:2020wbi,Tristram:2021tvh});  as a result, these data have become crucial in excluding the parameter space of inflationary models, therefore also excluding a wide range of these possible models.

In the models investigated, due to the simplicity of the potential, the only additional parameter included in the analysis is the dissipative ratio $Q_\star$, computed at horizon crossing; the other cosmological parameters are essentially the same as the $\Lambda$CDM model, namely the baryon and dark matter density parameters ($\Omega_b h^2$,$\Omega_c h^2$), the amplitude of the primordial scalar power spectrum $A_s$, the optical depth $\tau$, and the ratio between the sound horizon and the angular distance at decoupling $\theta$. Finally, the evidence of each model will be computed through the \texttt{MCEvidence} code \cite{Heavens:2017afc}, which can compute the Bayesian evidence of a given model from the generated MCMC chains \cite{Heavens:2017hkr}, while the mean values of the parameters and confidence contour plots are obtained through the \texttt{GetDist} code \cite{Lewis:2019xzd}.

\section{Results}\label{sec6}

\begin{table*}[t]
	\centering
	\begin{tabular}{>{\scriptsize}c >{\scriptsize}c >{\scriptsize}c >{\scriptsize}c >{\scriptsize}c }
        \hline
		\hline
		& Cubic coefficient & Linear coefficient & Inverse coefficient & $\propto H$ coefficient \\
		\hline
		{Parameter} & {mean}  & {mean}  & {mean} & {mean} \\
		\hline 
		\multicolumn{5}{c}{\texttt{PR4+BK18}}\\
		
		$\Omega_b h^2$   & $0.02230\pm 0.00012$ &  $0.02222\pm 0.00012$ & $0.02218\pm 0.00011$ & $0.02215\pm 0.00011$ \\
		$\Omega_{c} h^2$ & $0.11839\pm 0.00080$ &  $0.11964\pm 0.00087$ & $0.12021\pm 0.00077$ & $0.12042\pm 0.00084$ \\
		$100\theta$ & $1.04091\pm 0.00024$ &  $1.04077\pm 0.00025$ & $1.04071\pm 0.00024$ & $1.04073\pm 0.00026$  \\
		$\tau$ & $0.0608\pm 0.0059$ &  $0.0580\pm 0.0057$ & $0.0563\pm 0.0056$ & $0.0555\pm 0.0058$ \\
		$\ln(10^{10}A_s)$ & $3.049\pm 0.012$ &  $3.044\pm 0.011$ & $3.041\pm 0.012$ & $3.040\pm 0.011$ \\
		$\log_{10}Q_\star$ & $-2.32^{+0.50}_{-0.33}$  & $-2.11^{+0.80}_{-1.2}$ & $-1.9^{+1.1}_{-1.2}$ & $-1.4^{+1.1}_{-1.7}$ \\
        $H_{0}^{\ast}$ [Km/s/Mpc] & $67.87\pm 0.37$   & $67.31^{+0.37}_{-0.41}$ & $67.06\pm 0.35$ & $66.98\pm 0.37$ \\
		\hline
		\hline
        $\chi^2_{min}$ & $31124.3$ & $31124.7$ & $31126.2$ & $31127.3$\\
        $\ln{B}$ & $-4.133$ & $-0.779$ & $-1.656$ & $-2.187$ \\
		\hline
	\end{tabular}
	\caption{The estimates at $68\%$ confidence level (C.L.) for the cosmological parameters, when \textit{Planck} TTTEEE+lensing+BK18 data is considered for the quartic potential model. From left to right, the columns show the results for the cubic, linear, inverse and $\propto H$ dissipation coefficients. The $^\ast$ indicates a derived parameter. At the bottom, we show the values for the $\chi^2$ and Bayes factor estimates, when compared to Starobinsky inflation.}
	\label{tab:1}
\end{table*}

\begin{figure*}[t]
\centering
\includegraphics[width=\columnwidth]{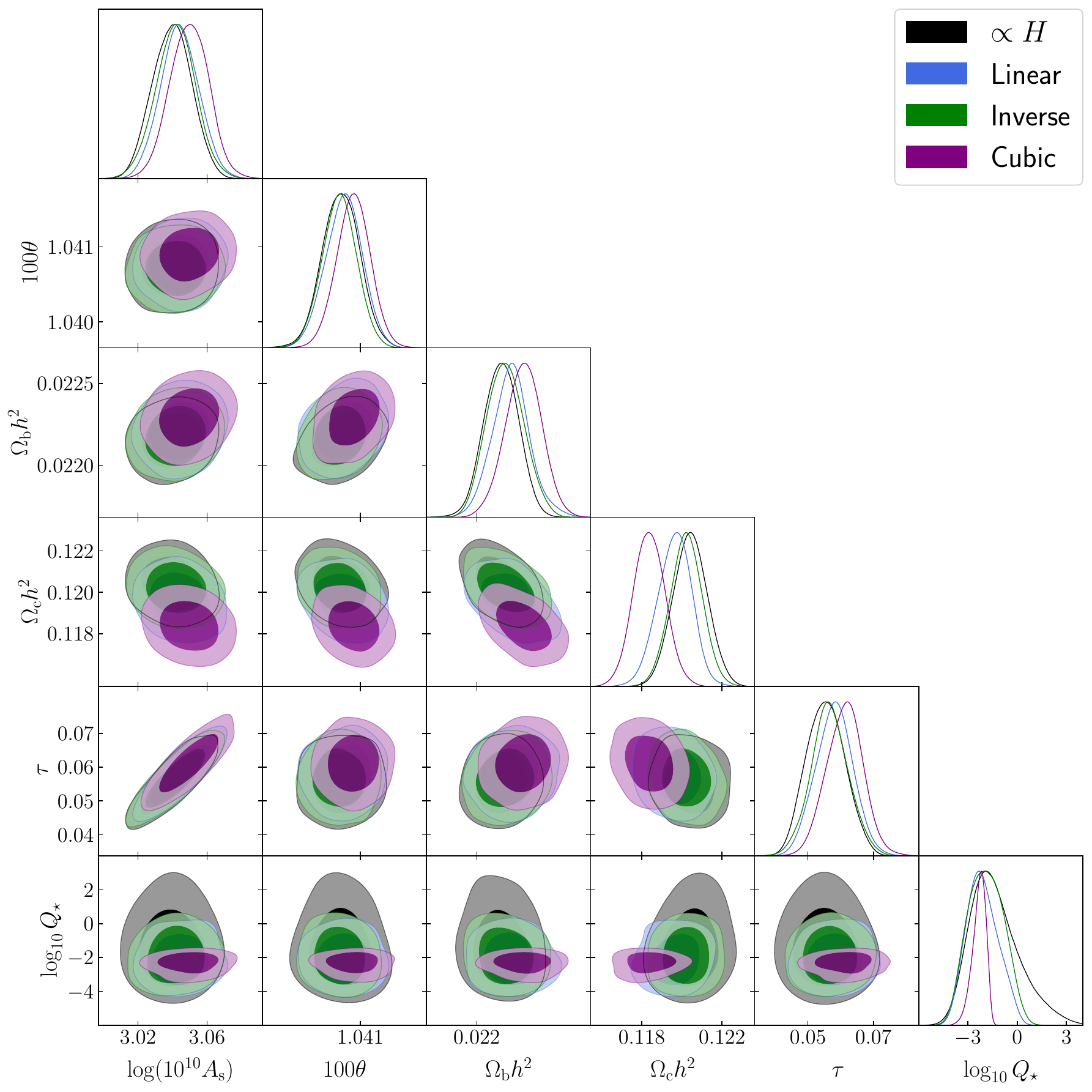}
\caption{Normalized posteriors and confidence contours for the WI models earlier considering the PR4+BK18 data sets.}
\label{fig:2}
\end{figure*}

The results of our analyses are shown in Table \ref{tab:1} and in fig. \ref{fig:2}. For the cubic dissipation model ($\Upsilon\propto T^3$), we found $\log_{10}Q_\star=-2.32^{+0.50}_{-0.33}$, with central value corresponding to $Q_\star=0.0045$, well into the weak dissipative regime. From fig. \ref{fig:1}, we note that such value is compatible with the current constraints on the spectral index, $n_s=0.9685$. It is noticeable how our estimate on $\log_{10}Q_\star$ is consistent with the one found in \cite{Arya:2017zlb}. As for the other cosmological parameters, our estimates are consistent with the standard model.

For the model with linear dissipation ($\Upsilon\propto T$), the results are quite similar. As the model allows for a higher dissipative level in agreement with the $n_s$ constraints, when compared to the cubic one, we expect that the estimates will reflect that. We find $\log_{10}Q_\star=-2.11^{+0.80}_{-1.2}$, with significantly higher uncertainty with respect to the cubic model. The $n_s-r$ limits allow the beginning of a strong regime ($Q\sim 1$) at $2\sigma$ level, but the weak dissipative regime ends up becoming more favorable as a better description, as we have $Q_\star = 0.008$. 

For the inverse dissipative model ($\Upsilon\propto T^{-1}$), we obtain $\log_{10}Q_\star=-1.9^{+1.1}_{-1.2}$. A general result is that the weak dissipative regime is favored, even if the model allows for a higher $Q_\star$. Finally, for a model where $\Upsilon\propto H$, which results in a constant $Q$, we find $\log_{10}Q_\star=-1.4^{+1.1}_{-1.7}$, again consistent with the weak regime, but still being the highest $Q_\star$ estimated, of around $Q_\star\simeq 0.04$.

The Bayes' factors for each model are displayed in Table \ref{tab:1}. Taking the approach of \cite{Planck:2018jri}, we compute $\ln B$ for each coefficient in comparison with the predictions of standard cold Starobinsky inflation, since, as detailed in the \textit{Planck} analysis, it provides the best fit to their data among the models considered there. According to Jeffreys' scale, there is an apparent evidence against the quartic warm inflation with a cubic dissipation coefficient. On the other hand, we notice an inconclusive result for the model with linear dissipation with respect to Starobinsky inflation. Finally, for the models given by the inverse and $\propto H$ dissipation coefficients, we find a weak evidence against these models in comparison with the reference one, models which also provide the worst fits to data, as seen by their $\chi^2$ values.

\begin{figure}[t]
\centering
\includegraphics[width=0.6\columnwidth]{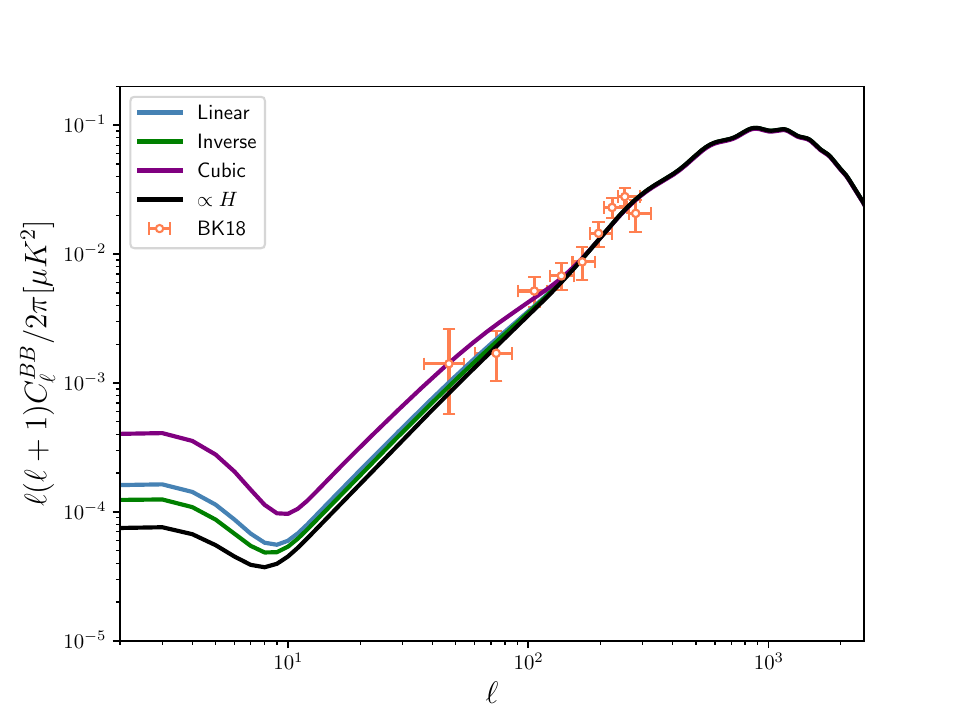}
\caption{CMB B-mode power spectra for the WI models investigated in this work. The data points corresponds to the \textit{BICEP/Keck Array} (BK18) estimates, also used in our numerical analysis.}
\label{fig:3}
\end{figure}

Since the tensor-to-scalar ratio always tends to very low values in warm inflation, we suspect that this result is a consequence of a better fit to the tensor power spectrum, since BK18 data is considered. We illustrate this in fig. \ref{fig:3}, where we show the best fit of the models to the B-mode polarization spectrum of the CMB. While all models show a good agreement with the data, it is noticeable how distinct they are at lower multipoles, reinforcing the importance of future data from surveys such as LiteBIRD \cite{Matsumura:2013aja,Hazumi:2019lys,LiteBIRD:2022cnt} designed to search for low-multipole E-modes and primordial B-modes. This is especially important for models such as warm inflationary ones, which predict a typically low tensor-to-scalar ratio, as such data will be able to start distinguishing classes of models, as well as different dissipative regimes.

\section{Conclusions}\label{sec7}

Although current CMB observations are consistent with an early inflationary epoch, knowing more about the exact mechanism that drove the accelerated expansion of the universe and the subsequent reheating era is necessary. In this context, the warm inflationary framework is an interesting idea seeking a more unified explanation of the early universe's dynamics that may also reconcile some of the simplest and best-motivated inflaton potentials with current CMB data.

In this work, we focused on the impact of the dissipation coefficient $\Upsilon$ that drives the decay of the inflaton field into radiation in the preference of a given model with data. By assuming the simple quartic potential, we tested four different functional forms for $\Upsilon$ against updated \textit{Planck} CMB likelihoods combined with the most recent data from the \textit{BICEP/Keck Array} collaboration, focused on the search for CMB B-modes. As each model furnishes a particular phenomenology, our results reflect these particularities, whether in the cosmological parameter estimates or its preference against a reference model. Our analysis found that the model given by a linear dependence of $\Upsilon$ on the temperature seems to be preferred, with an inconclusive Bayes factor compared to standard Starobinsky inflation.

Finally, it is worth mentioning that a crucial aspect of this kind of study is the computation of the primordial power
spectra for warm inflationary models. This has been a
recent topic of research \cite{Ballesteros:2023dno,Montefalcone:2023pvh}, and its correct calculation has profound implications on observational predictions, from CMB constraints to post-inflationary phenomena, such as the generation of gravitational waves \cite{Arya:2022xzc,Arya:2023pod,Correa:2023whf}. This description will become even more important as upcoming data from future collaborations are expected to restrict the parameter space so that some of the current viable models will be excluded.

\section*{Acknowledgments}

FBMS acknowledges financial support from the Programa de Capacita\c{c}\~ao Institucional do Observat\'orio Nacional (PCI/ON/MCTI). RdS is supported by the Coordena\c{c}\~ao de Aperfei\c{c}oamento de Pessoal de N\'ivel Superior (CAPES). JSA is supported by CNPq grant No. 307683/2022-2 and Funda\c{c}\~ao de Amparo \`a Pesquisa do Estado do Rio de Janeiro (FAPERJ) grant No. 259610 (2021). We also acknowledge the use of \texttt{cobaya}, \texttt{CAMB}, \texttt{MCEvidence} and \texttt{GetDist} codes. This work was developed thanks to the use of the National Observatory Data Center (CPDON).

\textbf{Note:} While finishing this manuscript, two papers with related analyses were submitted to ArXiv. The first \cite{DAgostino:2024nni} considers warm inflation constrained by pulsar timing arrays for the recent NANOGrav collaboration data. In the second work \cite{Kumar:2024hju}, the authors developed a general method to compute the primordial power spectra by solving the full differential equations, while presenting results for the WI quartic potential model with linear and cubic dissipation coefficients.

\bibliography{references}

\end{document}